\documentclass[journal]{IEEEtran}

\usepackage{cite}
\usepackage{amsmath,amssymb,amsfonts}

\usepackage{algorithm} 

\usepackage{subfigure}
\usepackage{textcomp}
\usepackage{bm}
\usepackage{amsmath,amsfonts}
\usepackage[thmmarks]{ntheorem}
\usepackage{algorithmic}
\usepackage{array}
\usepackage[caption=false,font=normalsize,labelfont=sf,textfont=sf]{subfig}
\usepackage{textcomp}
\usepackage{stfloats}
\usepackage{url}
\usepackage{verbatim}
\usepackage{graphicx}
\usepackage{amsmath}
\usepackage{amssymb}
\usepackage{enumitem}
\usepackage{color,xcolor}
\usepackage{xpatch}

\usepackage{booktabs}
\usepackage{graphicx}
\usepackage{tabularx}
\usepackage{titlesec}

\makeatletter
\def\changeBibColor#1{%
	\ifin@\color{blue}\else\normalcolor\fi
}
\xpatchcmd\@bibitem
{\item}
{\changeBibColor{#1}\item}
{}{}

\makeatother

\def\BibTeX{{\rm B\kern-.05em{\sc i\kern-.025em b}\kern-.08em
    T\kern-.1667em\lower.7ex\hbox{E}\kern-.125emX}}
\usepackage{balance}
\begin{document}
\title{Delay-Aware Digital Twin Synchronization in Mobile Edge Networks With Semantic Communications}
\author{Bin Li, Haichen Cai, Lei Liu, and Zesong Fei,~\IEEEmembership{Senior Member,~IEEE}

\thanks{Bin Li and Haichen Cai are with the School of Computer Science, Nanjing University of Information Science and Technology, Nanjing 210044, China (e-mail: bin.li@nuist.edu.cn; 202212210054@nuist.edu.cn).}	
\thanks{Lei Liu is with the Guangzhou Institute of Technology, Xidian University, Guangzhou 510555, China (e-mail: tianjiaoliulei@163.com).}	
\thanks{Zesong Fei is with the School of Information and Electronics, Beijing Institute of Technology, Beijing 100081, China (e-mail: feizesong@bit.edu.cn).}
}
\maketitle
\begin{abstract}
    The synchronization of digital twins (DT) serves as the cornerstone for effective operation of the DT framework. However, the limitations of channel capacity can greatly affect the data transmission efficiency of wireless communication. Unlike traditional communication methods, semantic communication transmits the intended meanings of physical objects instead of raw data, effectively saving bandwidth resource and reducing DT synchronization latency. Hence, we are committed to integrating semantic communication into the DT synchronization framework within the mobile edge computing system, aiming to enhance the DT synchronization efficiency of user devices (UDs). Our goal is to minimize the average DT synchronization latency of all UDs by jointly optimizing the synchronization strategy, transmission power of UDs, and computational resource allocation for both UDs and base station. The formulated problem involves sequential decision-making across multiple coherent time slots. Furthermore, the mobility of UDs introduces uncertainties into the decision-making process. To solve this challenging optimization problem efficiently, we propose a soft actor-critic-based deep reinforcement learning algorithm to optimize synchronization strategy and resource allocation. Numerical results demonstrate that our proposed algorithm can reduce synchronization latency by up to 13.2\% and improve synchronization efficiency compared to other benchmark schemes.
\end{abstract}


\section{Introduction}
As an emerging technology in the era of 6G, digital twin (DT) is capable of real-time mapping of real physical entities and environments to virtual spaces, achieving communication, collaboration, and information sharing between the physical and virtual worlds \cite{10443270}, making it a promising technology for network resource management \cite{10384610,10244089}. Furthermore, DT can support powerful artificial intelligence technologies, applying intelligent learning algorithms in virtual spaces along with real-time data provided by DT can assist physical entities in making more accurate and timely decisions, reducing the resource consumption of user decisions and improving the efficiency of resource utilization \cite{9429703}.

The DT framework is primarily composed of three components: the physical space, the digital space, and the communication space \cite{10012285}. DT synchronization is the cornerstone of this framework, conducting simulation and control operations through the digital replica of the physical world. To guarantee real-time monitoring of physical entities and processes, DT requires the processing and modeling of real-time data from physical entities \cite{9583902}, ensuring real-time update and precise synchronization of information.

However, the construction of DT models requires not only real-time data but also substantial computational capabilities. User devices (UDs) can sense real-time data, but their computational resources are limited, making the implementation of DT challenging \cite{10384610}. Mobile edge computing (MEC), as a promising paradigm, can deploy computational resource near users, ensuring low-latency task execution and DT maintenance, effectively addressing the resource constraints on UDs \cite{10025677,9780389}. Consequently, UDs can transmit sensed real-time data to MEC servers for modeling and updating the DT of physical entity, ensuring the communication efficiency and energy-saving of DT services \cite{9583902}. However, the bandwidth resource and energy on UDs are limited, as well as transmitting large amounts of data from UDs to the base station (BS) also results in latency and energy consumption. Therefore, how to process and transmit the real-time data sensed by UDs to achieve efficient DT synchronization is a critical issue that urgently needs to be addressed.

Recently, semantic communication has been recognized as a promising intelligent communication method for 6G. In contrast to traditional communication methods, semantic communication is task-oriented, extracting and transmitting only information related to tasks \cite{Qin2021SemanticCP}. This emerging communication paradigm extracts semantic features from the source information, obtains semantic information, and transmits the required meaning of physical objects rather than raw data, providing reliable and robust wireless communication services \cite{9530497,9450827,9398576}. The key goal of semantic communication has shifted from ensuring the accuracy of transmitted data and signal waveforms to ensuring the accurate understanding and correct use of transmitted semantic information, while prioritizing the conveyance of intended meaning with minimal data \cite{AcademicLecture,10558819}. This mode of communication not only minimizes data redundancy through semantic extraction, improving data transmission efficiency, but also shows strong robustness under conditions of limited bandwidth and low signal-to-noise ratio, effectively alleviating the bottleneck of channel capacity limitations \cite{10419853}.

To reduce the latency of UDs transmitting synchronization data and enhance DT synchronization efficiency, we develop a semantic communication-based DT synchronization framework. This framework aims to transmit the semantic information of physical objects rather than raw sensed data, conveying the intended meaning with the minimal amount of data possible, thereby reducing communication overhead and effectively lowering synchronization latency. In contrast to traditional synchronization methods, our proposed framework requires semantic extraction of the raw sensed data collected by UDs before transmission, followed by the recovery of semantic information on the edge server. While transmitting more raw sensed data can improve synchronization accuracy, it also increases the time required for synchronization. Therefore, an efficient synchronization strategy is necessary to determine the optimal semantic extraction approach and find the best resource allocation for semantic extraction and recovery, adhering to constraints on energy consumption, synchronization latency, and computational resource. In conclusion, the contributions of this work are as follows:

\begin{itemize}
    \item We propose a semantic communication-based DT synchronization framework in MEC that considers UDs' mobility and data sensing process. To balance synchronization latency and accuracy, a semantic extraction factor is introduced. The DT synchronization problem is formulated as a latency minimization problem, subject to constraints on energy consumption, synchronization latency, and computational resource.

    \item The formulated problem is a time-varying optimization problem, where the mobility of UDs leads to dynamic changes in the wireless environment, making it challenging for traditional convex-based methods to solve. To tackle this challenge, an efficient deep reinforcement learning (DRL) algorithm based on soft actor-critic (SAC) is proposed by jointly optimizing the semantic extraction factor, transmission power of UDs, and computational resource allocation for both UDs and the edge server. 

    \item We analyze the complexity of the proposed SAC-based algorithm and demonstrate its convergence. By analyzing the numerical results, our proposed SAC-based algorithm demonstrates superior performance in reducing the synchronization latency of UDs, compared to other benchmark schemes.

\end{itemize}

The remainder of this paper is structured as follows. Section II surveys related work. Section III describes the system model. Section IV formulates the optimization problem. Section V presents the solution to the problem under consideration. Numerical results and conclusions are detailed in Sections VI and VII, respectively.

\section{Related Work}
In this section, we introduce the related work on DT and semantic communication.

The DT technology can effectively assist MEC systems in building network topologies and optimizing resource allocation \cite{10384610}. For example, Guo \textit{et al.} \cite{9976231} proposed a resource allocation method based on federated learning (FL) for device-to-device aided DT edge networks, aiming to enhance network performance and address privacy issues. Liu \textit{et al.} \cite{10021296} presented a DT based unmanned aerial vehicle (UAV) assisted MEC network and utilized multi-agent collaboration to minimize the weighted energy consumption of users and UAVs. Hazarika \textit{et al.} \cite{10234627} introduced a DT-driven UAV-assisted internet of vehicles resource allocation algorithm to reduce energy consumption and network latency. He \textit{et al.} \cite{9887906} proposed a DT-assisted FL framework for resource allocation in heterogeneous cellular networks, which reduces transmission delay and protects user privacy. Zhang \textit{et al.} \cite{10335637} proposed a multi-FL service framework in DT-assisted MEC networks that optimizes device scheduling and resource allocation to address resource dynamics and mobile users, aiming to maximize utility across FL services. Du \textit{et al.} \cite{10558825} presented a distributed training architecture for mobile edge networks with semantic communications to address the computation bottleneck.

Semantic communication reduces communication overhead and significantly improves data transmission efficiency by transmitting and executing semantic information of raw data rather than processing the raw data of physical entities. Recently, Wang \textit{et al.} \cite{10419853} investigated the intelligent resource allocation issue in UAV-assisted spectrum sharing semantic communication networks to maximize the semantic spectrum efficiency of the auxiliary network. Wang \textit{et al.} \cite{10419536} proposed a UAV-assisted semantic communication system that jointly optimizes the trajectory of UAV, the number of semantic symbols, and transmit power by the four-SAC algorithm, achieving a balance between data transmission efficiency and energy efficiency. Ng \textit{et al.} \cite{10570867} introduced a stochastic semantic transmission scheme based on stochastic integer programming to address resource allocation amid stochastic user demand. Tang \textit{et al.} \cite{10530992} proposed a DT synchronization framework based on tiny machine learning, utilizing semantic communication in a UAV-assisted MEC to minimize synchronization latency, while considering the stability constraints of the virtual energy deficit queue. Yan \textit{et al.} \cite{10001594} investigated the quality-of-experience (QoE) aware resource allocation issue in semantic communication networks where an approximate measure of semantic entropy is introduced to quantify the semantic information for different tasks. 

The aforementioned studies are dedicated to exploring efficient DT frameworks in MEC systems or optimizing networks that incorporate semantic communication. However, to our knowledge, most existing studies have not taken the process of data sensing into consideration. This paper integrates semantic communication into DT synchronization to enhance data transmission efficiency, considering the uncertainties brought by data sensing process and mobility of UDs, and further investigates the process that combines sensing and data synchronization.

\section{System Model And Problem Formulation}
\begin{figure}[t]
    \centering
    \includegraphics[width=0.5\textwidth]{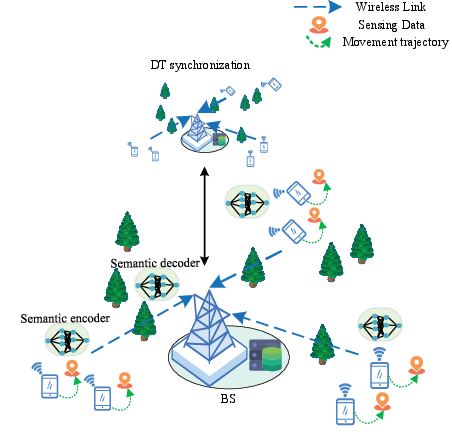}
    \caption{DT synchronization framework based on semantic communications.}
	\label{fig1}
\end{figure}

As shown in Fig. \ref{fig1}, we consider a DT synchronization system, which consists of $K$ UDs and one BS integrated with an edge server. UDs are equipped with sensors and cameras to sense various types of data from the surrounding environment. We define $k\in \mathcal{K}=\left\{ 1,2,\cdots ,K \right\}$ as the set of UDs.

To simplify, the total cycle of all UDs $T$ is equally divided into $N$ time slots with the length of $\tau =T/N$. The set of time slots is defined as $n\in \mathcal{N}=\left\{ 1,2,\cdots ,N \right\}$.
Table I presents all the important symbols defined in this paper.

\subsection{Mobility Model of UDs}
In this paper, we assume that the position of BS is fixed, while the UDs are randomly distributed within the target area in the first time slot. The positions of UDs only change between time slots, remaining constant within a time slot. We use ${{\textbf{p}}_{\text{B}}}=\left[ x_{\text{B}}^{{}},y_{\text{B}}^{{}},{{H}_{\text{B}}} \right]$ and $\textbf{p}_{k}^{{}}\left[ n \right]=\left[ x_{k}^{{}}\left[ n \right],y_{k}^{{}}\left[ n \right],0 \right]$ to denote the positions of BS and UD $k$ in time slot $n$, respectively. According to the Gauss-Markov random model described in \cite{9795902}, the coordinate of UD $k$ is updated as follows
\begin{align}
    {{x}_{k}}\left[n+1\right]={{x}_{k}}\left[n\right]+{{v}_{k}}\left[n\right]\cos ({{\theta }_{k}}\left[n\right])\tau ,   \\
    {{y}_{k}}\left[n+1\right]={{y}_{k}}\left[n\right]+{{v}_{k}}\left[n\right]\cos ({{\theta }_{k}}\left[n\right])\tau ,
\end{align}
where $v_{k}^{{}}\left[ n \right]$ and $\theta _{k}^{{}}\left[ n \right]$ denote the movement velocity and direction of UD $k$ in time slot $n$, respectively.

\subsection{Sensing Model}
In this paper, we adopt a ``sense-then-offload" \cite{9632276} mode where each UD first senses data from the target environment and then processes the sensed data. All UDs must complete semantic extraction, semantic information transmission, and semantic information recovery within a frame duration of $\left(1-\eta \right)\tau$ to meet real-time requirements. Here, $\eta \in \left[0,1\right]$ represents the time allocation factor for dividing the data sensing and processing stages of UDs. In diverse application scenarios, different types of data require different sensors for sensing. Various sensors collect data at different sampling rates, where sampling rate refers to the number of samples obtained per second. Therefore, the rate of data sensing is equal to the sample size multiplied by the sampling rate, and different sensors collect data at different rates. UDs are equipped with environmental sensors such as humidity, temperature and noise level, which provide data sensing capabilities for UDs. In this scenario, the data collected by UD $k$ during the sensing process can be denoted as
\begin{equation}
    D_{k}^{\text{s}}=\eta \tau v_{k}^{\text{s}},
\end{equation} 
where $v_{k}^{\text{s}}$ represents the sensing rate of UD $k$. The energy consumed by UD $k$ during the sensing process is represented as 
\begin{equation}
    E_{k}^{\text{s}}=p_{k}^{\text{s}}D_{k}^{\text{s}},
\end{equation}
where $p_{k}^{\text{s}}$ is the energy consumed for sensing one unit of bit data.

\begin{table}[t]
    \centering
    \captionsetup{labelfont={color=blue}}
    \caption{IMPORTANT NOTATIONS OF OUR ARTICLE}
    \setlength{\extrarowheight}{2pt} %
    \begin{tabularx}{\columnwidth}{p{0.1\columnwidth} X}
    \toprule[1pt] %
    Symbol & Meaning \\
    \midrule
    \midrule %
    $K$ & the number of UDs\\
    $N$ & the number of time slots\\
    $\tau$ & the duration of a time slot\\
    ${\textbf{p}}_{\text{B}}$ & the BS's position\\
    $\textbf{p}_{k}^{{}}\left[ n \right]$ & the $k$-th UD's position during time slot $n$\\
    $\eta$ & the time allocation factor for dividing the data sensing and processing stage\\
    $v_{k}^{\text{s}}$ & the sensing rate of UD $k$\\
    $p_{k}^{\text{s}}$ & the energy consumed for sensing per bit of data\\
    ${{D}_{k}}\left[ n \right]$ &the size of required sensed data\\
    $t_{k}^{\text{s}}\left[n\right]$ & the latency for sensing data of UD $k$ in the $n$-th time slot\\
    $e_{k}^{\text{s}}\left[n\right]$ & the energy consumption for sensing data of UD $k$ during time slot $n$\\
    ${{\varphi }_{k}}\left[n\right]$ & the semantic extraction factor of UD $k$ in time slot $n$\\
    ${{\lambda }_{k}}\left[n\right]$ & the computational demands for semantic extraction of UD $k$ in time slot $n$\\
    $t_{k}^{\text{en}}\left[n\right]$ & the computational latency for semantic extraction of UD $k$ in time slot $n$\\
    $C$ & the number of CPU cycles required to process unit bit of data\\
    ${{f}_{k}}\left[n\right]$ & the computational frequency of UD $k$ in time slot $n$\\
    $e_{k}^{\text{en}}\left[n\right]$ & The energy consumption for semantic extraction of UD $k$ in the $n$-th time slot\\
    $B$ & the total system bandwidth\\
    $p_{k}^{{}}\left[n\right]$ & the transmission power of UD $k$ during time slot $n$\\
    $u_{k}^{{}}\left[n\right]$ & the achievable rate of UD $k$ transmitting semantic information to BS in time slot $n$\\
    $t_{k}^{\text{up}}\left[n\right]$ & the UD $k$’s transmission latency during time slot $n$\\
    $e_{k}^{\text{up}}\left[n\right]$ & the energy consumption of UD $k$ for transmitting in the $n$-th time slot\\
    $t_{k}^{\text{de}}\left[n\right]$ & the recovery latency of UD $k$ in time slot $n$\\
    $f_{k}^{\text{e}}\left[n\right]$ & the computational resources allocated by the edge server to UD $ k$ during time slot $n$\\
    $t_{k}^{\text{dt}}\left[n\right]$ & the synchronization latency of UD $k$ in time slot $n$\\
    ${{t}_{k}}\left[n\right]$ & the total DT synchronization latency for UD $k$ in $n$-th time slot\\
    ${{e}_{k}}\left[n\right]$ & the total energy consumption of UD $k$ during time slot $n$\\
    $s^t$ & the DRL state in time slot $t$\\
    $a^t$ & the DRL action in time slot $t$\\
    $r^t$ & the DRL reward in time slot $t$\\

    \bottomrule[1pt] 
\end{tabularx}
\end{table}

\subsection{UD Synchronization Model}
In DT synchronization integrated with semantic communication, UDs are required to first extract the raw sensed data through semantic encoders, and then transmit the extracted semantic information. After the semantic information is transmitted to the edge server, the edge server needs to recover the semantic information through semantic decoders. Consequently, the extraction and recovery of semantic information impose additional computational workload on both UDs and the BS.

At the beginning of time slot $n$, UD $k$ needs to gather ${{D}_{k}}\left[ n \right]\in \left[ {{D}_{\min }},{{D}_{\max }} \right]$ bits of sensed data for DT synchronization, where ${D}_{\min }$ and ${D}_{\max }$ are the minimum and maximum amounts of sensed data required for DT synchronization, respectively. Then, the latency for sensing data of UD $k$ is calculated as 
\begin{equation}
    t_{k}^{\text{s}}\left[n\right]={{D}_{k}}\left[n\right]/v_{k}^{\text{s}},
\end{equation}
and the energy consumption for sensing date of UD $k$ can be computed as
\begin{equation}
    e_{k}^{\text{s}}\left[n\right]=p_{k}^{\text{s}}{{D}_{k}}\left[n\right].
\end{equation}

After collecting sensed data, UDs need to perform semantic extraction on the raw sensed data, which imposes additional computational loads on UDs. Let ${{\varphi }_{\min }}\le {{\varphi }_{k}}\left[n\right]\le 1$ represents the semantic extraction factor required by UD $k$ to extract the meaning of raw sensed data in time slot $n$, where ${\varphi }_{\min }$ denotes the minimum extraction factor that ensures the accuracy of semantic recovery at the edge server. The computational demands for semantic extraction of UD $k$ in time slot $n$ can be expressed as
\begin{equation}
    {{\lambda }_{k}}\left[n\right]=\frac{{{D}_{k}}\left[n\right]}{{{\left({{\varphi }_{k}}\left[n\right]\right)}^{x}}},\forall k\in \mathcal{K},n\in \mathcal{N},
\end{equation}
where $x$ is a constant and different types of DT synchronization tasks correspond to different $x$. The computational latency for semantic extraction of UD $k$ is given by
\begin{equation}
    t_{k}^{\text{en}}\left[n\right]=\frac{C{{\lambda }_{k}}\left[n\right]}{{{f}_{k}}\left[n\right]},\forall k\in \mathcal{K},n\in \mathcal{N},
\end{equation}
where $C$ is the number of CPU cycles required to process unit bit of data and ${{f}_{k}}\left[n\right]$ is the computational frequency of UD $k$ in time slot $n$ satisfying $0\le {{f}_{k}}\left[n\right]\le f_{k}^{\max }$, where $f_{k}^{\max }$ is the maximum available computational frequency of UD $k$. The energy consumption for semantic extraction of UD $k$ can be expressed as
\begin{equation}
    e_{k}^{\text{en}}\left[n\right]=C{{{k}_{\text{loc}}}}{{\lambda }_{k}}\left[n\right]{{\left({{f}_{k}}\left[n\right]\right)}^{2}},\forall k\in \mathcal{K},n\in \mathcal{N},
\end{equation}
where ${{k}_{\text{loc}}}$ is the effective capacitance coefficient of UD $k$.
\subsection{ Edge Synchronization Model}
DT needs to generate virtual replicas of physical entities in the real world on the edge server of BS, so the extracted semantic information needs to be uploaded to BS. Since the line-of-sight paths between UDs and BS are blocked by obstacles as depicted in Fig. 1, we model the channels from UDs to BS as Rayleigh fading channels. Then, the channel from UD $k$ to BS in time slot $n$ is expressed as follows
\begin{equation} 
    g_{k}^{\text{B}}\left[n\right]=\sqrt{\beta _{k}^{\text{B}}\left[n\right]}g_{k,\text{B}}^{\text{NLoS}}\left[n\right],
\end{equation}
where $\beta _{k}^{\text{B}}\left[n\right]=\beta _{0}^{{}}\left(d_{k}^{\text{B}}\left[n\right]\right)_{{}}^{\alpha _{3}^{{}}}$, $d_{k}^{\text{B}}\left[ n \right]=\left\| {\textbf{p}_{k}^{{}}\left[ n \right]}-{{\textbf{p}}_{\text{B}}} \right\|_{2}^{}$, ${\alpha}_{3}^{{}}$ denotes the path loss exponent, and $g_{k,\text{B}}^{\text{NLoS}}\left[ n \right]\sim \mathcal{C}\mathcal{N}\left( 0,1 \right)$.

By utilizing the orthogonal frequency division multiple access protocol, multiple UDs can transmit semantic information to BS simultaneously while avoiding signal interference among UDs. The achievable rate of UD $k$ transmitting semantic information to BS in time slot $n$ is calculated by
\begin{equation}
    u_{k}^{{}}\left[n\right]=B_{k}^{{}}\log _{2}^{{}}\left( 1+\frac{p_{k}^{{}}\left[n\right]\left| g_{k}^{\text{B}}\left[n\right] \right|_{{}}^{2}}{\sigma _{{}}^{2}} \right),
\end{equation}
where ${{B}_{k}}=B/K$ is the bandwidth allocated to UD $k$, $B$ is the total system bandwidth, $p_{k}^{{}}\left[n\right]$ is the transmission power of UD $k$, and $\sigma _{{}}^{2}$ is the power of Gaussian white noise.

The latency of transmitting semantic information from UD $k$ to BS can be computed by
\begin{equation}
    t_{k}^{\text{up}}\left[n\right]={{D}_{k}}\left[n\right]{{\varphi }_{k}}\left[n\right]u_{k}^{-1}\left[n\right],\forall k\in \mathcal{K},n\in \mathcal{N},
\end{equation}
and the energy consumption of UD $k$ for transmitting semantic information is given by
\begin{equation}
    e_{k}^{\text{up}}\left[n\right]=t_{k}^{\text{up}}{{p}_{k}}\left[n\right].
\end{equation}

After transmitting semantic information from UDs to BS, it is necessary to recover the semantic information into the original sensed data at the edge server. The recovery latency of the semantic information can be calculated by
\begin{equation}
    t_{k}^{\text{de}}\left[n\right]=\frac{C{{D}_{k}}\left[n\right]{{\varphi }_{k}}\left[n\right]}{{{\left({{\varphi }_{k}}\left[n\right]\right)}^{y}}f_{k}^{\text{e}}\left[n\right]},
\end{equation}
where $y$ is a constant and different types of DT synchronization tasks correspond to different $y$, $f_{k}^{\text{e}}\left[n\right]$ is the CPU frequency assigned by the edge server for the recovery of semantic information uploaded by UD $k$ which needs to satisfy $\sum\limits_{k\in \mathcal{K}}{f_{k}^{\text{e}}\left[n\right]}\le f_{\text{e}}^{\max }$, with $f_{\text{e}}^{\max }$ denoting the maximum available CPU frequency of the edge server. Since the edge server is continuously powered, the energy consumption for semantic recovery can be neglected.

In the DT synchronization framework within edge computing systems, synchronization latency is a crucial metric to assess the performance of DT applications. Therefore, the synchronization latency is considered as the optimization goal of the synchronization strategy. The synchronization latency of UD $k$ which includes semantic extraction latency $t_{k}^{\text{en}}\left[n\right]$, semantic information transmission latency $t_{k}^{\text{up}}\left[n\right]$, and semantic recovery latency $t_{k}^{\text{de}}\left[n\right]$, can be expressed as 
\begin{equation}
   t_{k}^{\text{dt}}\left[n\right]=t_{k}^{\text{en}}\left[n\right]+t_{k}^{\text{up}}\left[n\right]+t_{k}^{\text{de}}\left[n\right].
\end{equation}
The synchronization task corresponding to the sensed data uploaded by UD $k$ needs to be processed and completed before the deadline. Consequently, the following constraint must be satisfied
\begin{equation}
    t_{k}^{\text{dt}}\left[n\right]\le \left(1-\eta\right) \tau ,\forall k\in \mathcal{K},n\in \mathcal{N}.
\end{equation}

 The total DT synchronization latency for UD $k$ consists of two parts: data sensing latency $t_{k}^{\text{s}}\left[n\right]$ and synchronization latency $t_{k}^{\text{dt}}\left[n\right]$, which is given by
 \begin{equation}
    {{t}_{k}}\left[n\right]=t_{k}^{\text{s}}\left[n\right]+t_{k}^{\text{dt}}\left[n\right].
\end{equation}
 The total energy consumption of UD $k$ during synchronization is expressed as 
\begin{equation}
    {{e}_{k}}\left[n\right]=e_{k}^{\text{s}}\left[n\right]+e_{k}^{\text{en}}\left[n\right]+e_{k}^{\text{up}}\left[n\right].
\end{equation}

\subsection{Problem Formulation}
In this paper, our goal is to minimize the average synchronization latency of all UDs by jointly optimizing the semantic extraction factor, transmission power, and resource allocation, subject to constraints on energy consumption, synchronization latency, and computational resource. The average synchronization latency minimization problem can be formulated as
\begin{subequations}
    \begin{align}
        &\underset{\varphi \left[n\right],f\left[n\right],{{f}_{\text{e}}}\left[n\right],p\left[n\right]}{\mathop{\min }}\,\frac{1}{N}\sum\limits_{n=1}^{N}{T\left[n\right]}\\
        \text{s.t.}~ 
        &t_{k}^{\text{s}}\left[n\right]\le \eta \tau ,\forall k\in \mathcal{K},n\in \mathcal{N},\\
        &t_{k}^{\text{dt}}\left[n\right]\le \left(1-\eta \right) \tau ,\forall k\in \mathcal{K},n\in \mathcal{N},\\
        &{{E}_{k}}\left[n\right]\le E_{\text{u}}^{\max },\forall k\in \mathcal{K},\\
        &{{\varphi }_{\min }}\le {{\varphi }_{k}}\left[n\right]\le 1,\forall k\in \mathcal{K},n\in \mathcal{N},\\
        &0\le {{f}_{k}}\left[n\right]\le f_{\text{u}}^{\max },\forall k\in \mathcal{K},n\in \mathcal{N},\\
        &p_{\text{u}}^{\min }\le {{p}_{k}}\left[n\right]\le p_{\text{u}}^{\max },\forall k\in \mathcal{K},n\in \mathcal{N},\\
        &0\le \sum\limits_{k\in \mathcal{K}}{f_{k}^{\text{e}}\left[n\right]}\le f_{\text{e}}^{\max },\forall n\in \mathcal{N},
    \end{align}
\end{subequations}
where $T\left[n\right]=\sum\nolimits_{k\in \mathcal{K}}{{{t}_{k}}\left[n\right]}$. Constraints (19b) and (19c) guarantee that data sensing and synchronization are completed within the slot duration $\tau $ to meet real-time requirements. Constraint (19d) limits the energy consumption of UD $k$ in time slot $n$. Constraint (19e) ensures the DT synchronization accuracy. Constraints (19f) and (19h) represent the computational resource limitations of UDs and edge server, respectively. Constraint (19g) indicates the transmission power limitation of UDs.

The nonlinear objective function and the non-convexity of constraints make problem (19a) a non-convex optimization problem as they involve the multiplicative among the variables. Also, the mobility of UDs introduces uncertainties in solving the problem. Based on the above analysis, traditional convex optimization methods are inadequate to effectively solve this problem. Therefore, we adopt an SAC-based DRL algorithm to solve this problem.

\section{problem solution}
In this section, we proposed an SAC-based algorithm to optimize the semantic extraction factor, transmission power and allocation of computational resource for both UDs and the edge server. The Markov decision process (MDP) model and the SAC-based algorithm are detailed in the following.
\subsection{Modeling of MDP}
We model the problem of minimizing the synchronization latency in MEC system as an MDP. The MDP consists of three elements: state, action, and the reward, which are defined as follows.

{\bf{1) State:}} The environment state in time slot $n$, i.e., $S\left[n\right]$ includes two parts: i) the distance between UDs and BS, i.e., ${{d}_{\text{b}}}\left[n\right]=\left\{d_{k}^{\text{b}}\left[n\right],\forall k\in \mathcal{K}\right\}$; ii) the amount of data that UDs need to sense for DT synchronization, i.e., $D\left[n\right]=\left\{{{D}_{k}}\left[n\right],\forall k\in \mathcal{K}\right\}$, denoted as 
\begin{equation}
    S\left[n\right]=\left\{{{d}_{\text{b}}}\left[n\right],D\left[n\right]\right\}.
    \label{eq20}
\end{equation}

{\bf{2) Action:}} The action space of the formulated MDP in time slot $n$, i.e., $a\left[n\right]$ consists of the semantic extraction factor, transmission power and computational resource allocation for both UDs and the edge server, denoted as 
\begin{equation}
    \begin{aligned}
        a\left[n\right]=\left\{ {{\varphi }_{k}}\left[n\right],\ {{f}_{k}}\left[n\right],{{p}_{k}}\left[n\right],\ f_{k}^{\text{e}}\left[n\right] \right\}. \\ 
    \end{aligned}
    \label{eq21}
\end{equation}

{\bf{3) Reward:}}  DRL aims to maximize the reward function, but our optimization goal is to minimize the average synchronization latency. In addition, the reward function needs to include penalties for failing to satisfy the constraints. Based on the above analysis, we need to sum the objective function and the penalties and then take the inverse. Therefore, the reward function is defined as follows

\begin{equation}
    r\left[n\right]=-T\left[n\right]-P,
\label{eq22}
\end{equation}
where $P={{P}_{\text{t}}}+{{P}_{\text{e}}}+{{P}_{\text{f}}}$ denotes the sum of penalties, ${{P}_{\text{t}}}=\left( \sum\nolimits_{k\in \mathcal{K}}{t_{k}^\text{dt}\left[n\right]-(1-\eta )\tau } \right)W$ is the penalty set for DT synchronization timeout, ${{P}_{\text{e}}}=\left( \sum\nolimits_{k\in \mathcal{K}}{{{E}_{k}}\left[n\right]-E_\text{u}^{\max }} \right)W$ is the penalty set for the energy consumption of UDs exceeding the budget, ${{P}_{\text{f}}}=\left( \sum\nolimits_{k\in \mathcal{K}}{f_{k}^{\text{e}}\left[n\right]}-f_{\text{e}}^{\max } \right)W$ is the penalty set for the computational resource of edge server exceeding the limit, with $W$ a positive constant.

\subsection{SAC-Based Algorithm for DT Synchronization System}
SAC is an off-policy DRL algorithm that differs from traditional DRL algorithms in several key ways. Unlike methods that focus solely on maximizing cumulative rewards, SAC also takes the entropy of the policy into consideration, viewing it as an important objective. The goal of the SAC algorithm is to maximize a weighted sum of cumulative rewards and policy entropy. This approach encourages the agent to introduce more randomness in its action selection, which can be beneficial in finding the optimal solution. By promoting a more stochastic policy, SAC aims to explore a broader range of actions, potentially leading to better long-term performance and a more robust policy. Thus, we opt for SAC algorithm to solve the problem. In the remaining of this section, we will introduce the workflow of SAC algorithm in details.

The SAC algorithm employs a maximum entropy framework that encourages exploration by maximizing the entropy of policy. This policy maximizes the expected reward and entropy, the optimal policy ${{\pi }^{*}}$ can be expressed as follows
\begin{equation}
    \begin{aligned}
        \pi^* =& \arg\max_{\pi} \, \mathbb{E}_{(s^t, a^t) \sim \rho_{\pi}} \\
        &\left[ \sum_{t=0}^{\infty} \gamma^t r\left(s^t, a^t\right) + \alpha H\left(\pi\left(\cdot | s^t\right)\right) \right], 
    \end{aligned}
    \label{eq23}
\end{equation}
where $\gamma $ is the discount factor, $\rho _{\pi }^{{}}$ denotes the state-action trajectory distribution following the policy $\pi $, $H\left(\pi \left(\cdot |{{s}^{t}}\right)\right)={{E}_{{{a}^{t}}}}\left[-{{\log }_{2}}\pi \left({{a}^{t}}|{{s}^{t}}\right)\right]$ is the entropy of policy, the temperature parameter $\alpha $ can balance the entropy against the reward.

The soft Q-function for a given policy $\pi $ based on the Bellman expectation equation is defined as
\begin{equation}
    Q_{\pi }^{{}}\left( s_{}^{{t}},a_{}^{{t}} \right)=r_{}^{{t}}+\gamma \mathbb{E}_{s_{}^{{t+1}}\sim p_{s}^{{}}}^{{}}v_{\pi }^{{}}\left( s_{}^{{t+1}} \right).
    \label{eq24}
\end{equation}

Similarly, the soft state-value function takes the entropy into consideration, which is given by
\begin{equation}
    v_{\pi }^{{}}\left( s_{}^{{t}} \right)=\mathbb{E}_{a_{}^{{t}}\sim\pi }^{{}}\left[ Q_{\pi }^{{}}\left( s_{}^{{t}},a_{}^{{t}} \right) \right.-\left. \alpha \log _{2}^{{}}\left( \pi \left( a_{}^{{t}}\left| s_{}^{{t}} \right. \right) \right) \right].
    \label{eq25}
\end{equation}

The SAC algorithm utilizes neural networks to approximate the Q-function and policy, parameterizing the soft Q-function and policy as ${{Q}_{\theta }}\left(s,a\right) $ and ${{\pi }_{\phi }}\left(a|s\right) $, where $\theta $ and $\phi $ are parameters of the neural networks.
According to \cite{9420264}, the parameters of the soft Q-network are trained by minimizing the Bellman residual, which is given by
\begin{equation}
    \begin{aligned}
        {{L}_{Q}}\left(\theta \right)=&{{\mathbb{E}}_{({{s}^{t}},{{a}^{t}})\sim\mathcal{B}}} \\
        &\left[ \frac{1}{2}{{\left({{Q}_{\theta }}\left({{s}^{t}},{{a}^{t}}\right)-\hat{Q}\left({{s}^{t}},{{a}^{t}}\right)\right)}^{2}} \right],
    \end{aligned}          
    \label{eq26}
\end{equation}
where $\mathcal{B}$ is the replay buffer, $\hat{Q}\left({{s}^{t}},{{a}^{t}}\right)$ is the target soft Q-network.

Similarly, by minimizing the expected Kullback-Leibler divergence, parameters of the policy network can be trained
\begin{equation}
    \begin{aligned}
        {{L}_{\pi }}\left(\phi \right)=&{{\mathbb{E}}_{{{s}^{t}}\sim\mathcal{B}}}{{\mathbb{E}}_{{{a}^{t}}\sim{{\pi }_{\phi }}}} \\
        &\left[ \alpha \log {{\pi }_{\phi }}\left({{a}^{t}}|{{s}^{t}}\right)-{{Q}_{\theta }}\left({{s}^{t}},{{a}^{t}}\right) \right].
    \end{aligned}           
    \label{eq27}
\end{equation}

\begin{figure}[t]
    \centering
    \includegraphics[width=0.5\textwidth]{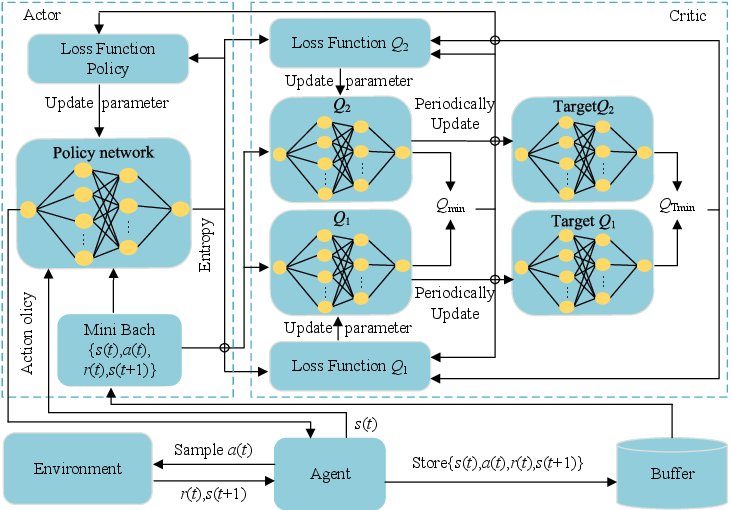}
    \caption{Training workflow of the SAC algorithm.}
	\label{fig2}
\end{figure}

During the training process, the variation in system rewards can cause the temperature parameter $\alpha$ to change dynamically, leading to instability in training. Therefore, using a fixed temperature is unreasonable. Thus, during the training process, the temperature parameter should be automatically adjusted while updating the network parameters. In this paper, the temperature parameter is adjusted by minimizing the loss function
\begin{equation}
    \begin{aligned}
        L\left(\alpha \right)={{E}_{s\sim\mathcal{B}}}{{E}_{a\sim\pi }}\left[-\alpha \log {{\pi }_{\phi }}\left(a|s\right)-\alpha {{H}_{0}}\right],
    \end{aligned}          
    \label{eq28}
\end{equation}
where ${{H}_{0}}$ is the target entropy value. As described in \cite{9681874}, the optimal dual variable at each time step is given by 
\begin{equation}
    \begin{aligned}
        \alpha _{t}^{*}=&\underset{\alpha _{{}}^{t}}{\mathop{\arg \min }}\,\mathbb{E}_{a_{{}}^{t}\sim\pi _{t}^{*}}^{{}} \\
        &\left[-\alpha _{{}}^{t}\log _{2}^{{}}\left(\pi _{t}^{*}\left({{a}^{t}}|{{s}^{t}};\alpha _{{}}^{t}\right)\right)-\alpha _{{}}^{t}{{H}_{0}}\right],
    \end{aligned}          
    \label{eq29}
\end{equation}
where $\pi _{t}^{*}\left({{a}^{t}}|{{s}^{t}};\alpha _{{}}^{t}\right)$ indicates the optimal policy corresponding to temperature $\alpha _{{}}^{t}$. Fig. \ref{fig2} shows the information flow of SAC and Algorithm 1 shows the detailed process of SAC-based algorithm.

\begin{algorithm}[t]
    \caption{SAC-Based Algorithm}
    \begin{algorithmic}[1]
    \STATE Initialize $\theta  _{i}^{{}}$, $\widehat{{{\theta }_{i}}}\left( i=1,2 \right)$, $\phi $, $\mathcal{B}$, lr, $\gamma $, and $H_{\min }^{{}}$.
    \FOR{each epoch}
        \FOR{ each step}
            \STATE Execute action based on current policy $a_{}^{{t}}\sim \pi _{\phi }^{{}}$;
            \STATE Calculate reward $r\left({{s}^{t}},{{a}^{t}}\right)$ via (22), and observe next state $s_{}^{{t+1}}$;
            \STATE Store transition $\left\{ s_{}^{{t}},a_{}^{{t}},r({{s}^{t}},{{a}^{t}}),s_{}^{{t+1}} \right\}$ in $\mathcal{B}$;
            \STATE Update the positions of UEs;
            \STATE Randomly sample a mini-batch transitions from $\mathcal{B}$.
            \STATE Update the critic network $\theta  _{i}^{{}}$ by minimizing Bellman residual (26);
            \STATE Update the actor network $\phi $ by minimizing expected Kullback-Leibler divergence (27);
            \STATE Update target network $\widehat{{{\theta }_{i}}}$ parameters periodically;
        \ENDFOR
    \ENDFOR
    \RETURN{The optimal policy $\pi _{\phi }^{*}$}.
\end{algorithmic}
\label{algorithm1}
\end{algorithm}
\subsection{Complexity Analysis}
The complexity of the proposed SAC-based algorithm mainly comes down to the iterative updating of neural network parameters in the DRL algorithm. The computational complexity of backpropagation and gradient descent for the actor networks and critic networks is $ O\left( {{n}_\text{epoch}}{{n}_\text{step}}\sum\nolimits_{i=1}^{M}{{{n}_{i}}{{n}_{i-1}}} \right)$, where $M$ is the number of layers of the neural network, and ${{n}_{i}}$ is the number of neurons in the $i$-th layer. ${n}_\text{epoch}$ and ${n}_\text{step}$ are the number of training epochs and the length of each epoch, respectively.
\begin{center}
    \begin{table}[t]
        \centering
        \caption{SIMULATION PARAMETERS}
        \begin{tabularx}{0.9\linewidth}{XX|XX}
            \hline
             {Parameters} &  {Values} &  {Parameters} &  {Values}\\
            \hline
            {$D_{m}^{\max }$} &{0.8 Mbit} & {$C$}& {300 cycles/bit}\\
            {$D_{m}^{\min }$} & {0.6 Mbit} & {${{\beta }_{0}}$}& {-30 dB}\\
            {$B$} & {0.2 MHz} &  {$\sigma^2$} & {-80 dBm}\\
           {$p_{\text{u}}^{\max }$} & {0.1 W} & {$E_{\text{u}}^{\max }$} & {0.5 J}\\
           {$p_{\text{u}}^{\min }$} & {0.01 W} & {$f_{\text{u}}^{\max}$} & {1 GHz}\\
           {${{k}_{\text{loc}}}$} & {${10}^{-27}$} & {$f_{\text{e}}^{\max}$}& {10 GHz}\\
            {$x$} & {1.2}  & {$y$} &{1.5}\\
            \hline
        \end{tabularx}
    \end{table}
    \label{tab2}
\end{center}

\begin{table}[t]
    \centering
    \caption{Parameter Settings in the SAC Algorithm}
    \begin{tabular}{cccc}\hline
    Parameter & Value\\\hline
    Learning rate lr & 0.0001\\
    Number of epoch ${n}_\text{epoch}$ & 20\\
    Number of step ${n}_\text{step}$ & 5000\\
    Mini-batch size & 256\\
    Replay memory size $\mathcal{B}$& 1000000\\
    Discount factor $\gamma $ & 0.99\\
    Number of hidden layers & 2\\
    Activation function & ReLU\\
    Optimizer & Adam\\
    Target network update frequency & 320\\
    Constant $W$ & 10\\\hline
    \end{tabular}
\label{tab3}
\end{table}

\section{NUMERICAL RESULTS}

In this section, we conduct numerical results of Python 3.7 with PyTorch to evaluate the performance of our proposed SAC-based algorithm. In this simulation experiment, mobile UDs are randomly distributed within a circle centered at (50, 0, 0) meters with a radius of 5 meters, while the location of BS is set at (0, 0, 0) meters \cite{10530992,10495829}. The number of UDs is set to $K=6$. The cycle and the number of time slots are $ T=30 $ s and $N=25$. The time allocation factor is set to $\eta=0.25$. The amount of sensed data required for DT synchronization is set to ${{D}_{m}}\in \left[ 0.6,0.8 \right]$ Mbit. The number of CPU cycles required for each bit of data computation is $C=300$ cycles/bit. The total bandwidth of the system is set to $B=0.2$ MHz. The channel power gain and power of Gaussian white noise are set as $\beta _{0}^{{}}=-30$ dB and $\sigma _{{}}^{2}=-80$ dBm. The maximum and minimum transmission powers of UDs are set to $p_{\text{u}}^{\max }=0.1$ W and $p_{\text{u}}^{\min }=0.01$ W. The energy consumption budget of UDs is set to $E_{\text{u}}^{\max }=0.5$ J. The maximum available computation frequency of UDs is set to $f_{\text{u}}^{\max}=1$ GHz. The maximum available CPU frequency provided by the edge server is set to $f_{\text{e}}^{\max}=10$ GHz. The effective capacitance coefficient of UDs is set to ${{k}_{\text{loc}}}=10_{{}}^{-27}$. Constants $x$ and $y$ corresponding to a specific DT synchronization task are set to $1.2$ and $1.5$. The confidence interval is set as 95\%. The parameter settings of simulation are summarized in Table II, according to prior works \cite{10530992,10495829}, and the DRL hyperparameters are listed in Table III.
To better evaluate the effectiveness of the proposed algorithm, we introduce three benchmark schemes for comparison:
\begin{itemize}
    \item {\bf{Proximal Policy Optimization (PPO)-Based Design}}: In this case, the synchronization strategy and resource allocation are optimized by the PPO-based algorithm. PPO is a popular and reliable advanced DRL algorithm, favored for its high performance in complex tasks.
\end{itemize}
\begin{itemize}
    \item {\bf{Without Semantic Communication}}: This scheme is labeled as ``w/o SC''. In this case, the semantic extraction factor is set to ${{\varphi }_{k}}[n]=1$. After sensing the data, UDs transmit the raw sensed data directly to the BS. 
\end{itemize}
\begin{itemize}
    \item {\bf{Random Synchronization Strategy}}: In this case, semantic extraction factors for each time slot is randomly generated. 
\end{itemize}

\begin{figure}[t]
    \centering
    \includegraphics[width=0.95\columnwidth]{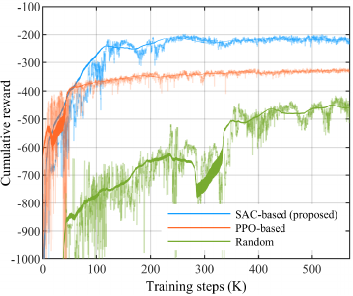}
    \caption{The convergence performance comparison.}
	\label{fig3}
\end{figure}

In Fig. \ref{fig3}, we compare the convergence performance of the proposed algorithm with two benchmark schemes. It can be observed from the figure that as training progresses, the rewards of three schemes gradually stabilize, and the reward of the proposed algorithm is higher than the other two benchmark schemes. This is because the SAC algorithm takes into account the entropy of policy in the learning objectives, and increasing the entropy value encourages the agent to explore more actions. Furthermore, it can be observed that the scheme with randomly generated semantic extraction factors achieves the lowest reward after convergence, which demonstrates the necessity and effectiveness of introducing semantic communication and optimizing synchronization strategy under adverse channel conditions.
\begin{figure}[t]
    \centering
    \includegraphics[width=0.95\columnwidth]{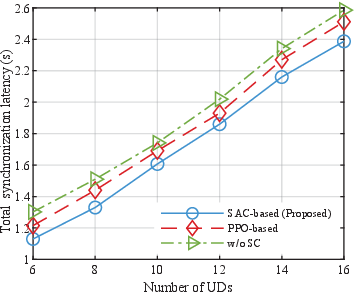}
    \caption{DT synchronization latency versus the number of UDs.}
	\label{fig4}
\end{figure}

Fig. \ref{fig4} shows the impact of the number of UDs $K$ on total DT synchronization latency. For all these schemes, we observe that the total DT synchronization latency increases significantly as the number of UDs $K$ increases. This is primarily due to two reasons: firstly, as the number of UDs grows, the bandwidth available to each UD decreases. Secondly, the computational resource of the edge server is finite, and as the number of UDs increases, the computational resources that can be allocated to each UD decrease. Consequently, as the number of UDs $K$ increases, the total DT synchronization latency also increases. This demonstrates the necessity of optimizing resource allocation in systems with limited bandwidth and computational resource.

\begin{figure}[t]
    \centering
    \includegraphics[width=0.95\columnwidth]{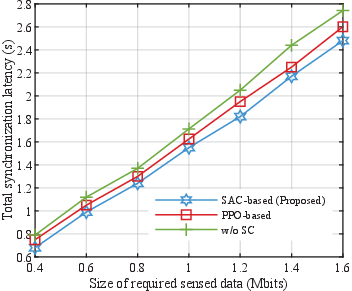}
    \caption{DT synchronization latency versus the size of required sensed data.}
	\label{fig5}
\end{figure}

Fig. \ref{fig5} compares the total DT synchronization latency of UDs under different schemes with varying amounts of required sensed data for synchronization. It is evident that the total DT synchronization latency for all schemes increases with the amount of required sensed data. This is due to the fact that, as the amount of sensed data increases, both the data sensing latency and synchronization latency for UDs also increase. Furthermore, the proposed SAC-based algorithm has the lowest total DT synchronization latency, while the scheme ``w/o SC'' has the highest. This indicates that the proposed SAC-based algorithm can effectively optimize the synchronization strategy and resource allocation, enabling UDs to complete DT synchronization more efficiently and thus reduce synchronization latency. This further demonstrates the importance of incorporating semantic communication.

\begin{figure}[t]
    \centering
    \includegraphics[width=0.95\columnwidth]{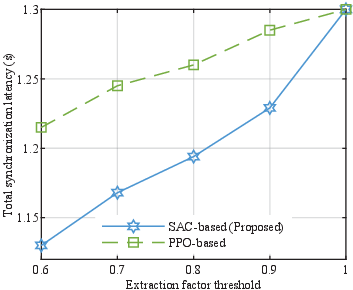}
    \caption{DT synchronization latency versus the extraction factor threshold.}
	\label{fig6}
\end{figure}

Fig. \ref{fig6} illustrates a comparison of total DT synchronization latency for UDs under different schemes with varying semantic extraction factor thresholds. When the semantic extraction factor threshold is set to ${{\varphi }_{k}}[n]=1$, the proposed SAC-based algorithm is equivalent to the second benchmark scheme. Therefore, in this comparison, we only contrast the proposed algorithm with the first benchmark scheme. Observing Fig. 6, it is obvious that as the semantic extraction factor threshold increases, the total DT synchronization latency correspondingly increases as well. This is because as the semantic factor threshold increases, the amount of data obtained after semantic extraction from the same amount of raw sensed data also increases, which means that UDs need to transmit more data to the BS. This demonstrates that semantic communication can effectively extract semantics from raw sensed data, reduce data redundancy, and enhance data transmission efficiency. Moreover, the synchronization latency of our proposed SAC-based algorithm is significantly lower than that of the PPO-based scheme, which indicates that the performance of our proposed algorithm is better than the PPO-based scheme.
\begin{figure}[t]
    \centering
    \includegraphics[width=0.95\columnwidth]{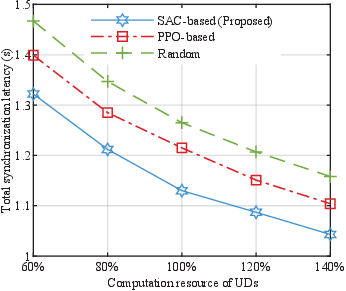}
    \caption{DT synchronization latency versus the computation resource of UDs.}
	\label{fig7}
\end{figure}

Fig. \ref{fig7} compares the total DT synchronization latency of UDs as it varies with the computational resource of UDs under different schemes. In this comparison, the algorithm we proposed is contrasted with the first and third benchmark schemes. It can be clearly observed that as the computational resource of the UDs increase, the total DT synchronization latency significantly decreases. This phenomenon is attributed to the accelerated processing speed during the semantic extraction process due to the enhancement of computational resource. Additionally, the proposed SAC-based algorithm performs the best in reducing the total synchronization latency of UDs, which further proves that our algorithm is more efficient in optimizing synchronization strategy and resource allocation, thereby achieving superior performance.

\section{CONCLUSION}
In this paper, we proposed a semantic communication-based DT synchronization framework in MEC systems that considers the mobility of UDs and data sensing processes. We jointly optimized the semantic extraction factor, transmission power of UDs and computational resource allocation for both UDs and the edge server to minimize the DT synchronization latency, and then developed the SAC-based algorithm to flexibly learn the suboptimal strategies. The performance of our proposed algorithm was verified by comparing numerical results with benchmark schemes. In our future work, we will research topics related to the accuracy of DT.

\bibliographystyle{IEEEtran}
\bibliography{IEEEabrv,refs}

\begin{thebibliography}{10}
\providecommand{\url}[1]{#1}
\csname url@samestyle\endcsname
\providecommand{\newblock}{\relax}
\providecommand{\bibinfo}[2]{#2}
\providecommand{\BIBentrySTDinterwordspacing}{\spaceskip=0pt\relax}
\providecommand{\BIBentryALTinterwordstretchfactor}{4}
\providecommand{\BIBentryALTinterwordspacing}{\spaceskip=\fontdimen2\font plus
\BIBentryALTinterwordstretchfactor\fontdimen3\font minus
  \fontdimen4\font\relax}
\providecommand{\BIBforeignlanguage}[2]{{%
\expandafter\ifx\csname l@#1\endcsname\relax
\typeout{** WARNING: IEEEtran.bst: No hyphenation pattern has been}%
\typeout{** loaded for the language `#1'. Using the pattern for}%
\typeout{** the default language instead.}%
\else
\language=\csname l@#1\endcsname
\fi
#2}}
\providecommand{\BIBdecl}{\relax}
\BIBdecl

\bibitem{10443270}
P.~Consul, I.~Budhiraja, D.~Garg, N.~Kumar, R.~Singh, and A.~S. Almogren, ``A
  hybrid task offloading and resource allocation approach for digital
  twin-empowered {UAV}-assisted {MEC} network using federated reinforcement
  learning for future wireless network,'' \emph{IEEE Transactions on Consumer
  Electronics}, vol.~70, no.~1, pp. 3120--3130, Feb. 2024.

\bibitem{10384610}
Y.~Dai, J.~Zhao, J.~Zhang, Y.~Zhang, and T.~Jiang, ``Federated deep
  reinforcement learning for task offloading in digital twin edge networks,''
  \emph{IEEE Transactions on Network Science and Engineering}, vol.~11, no.~3,
  pp. 2849--2863, May-Jun. 2024.

\bibitem{10244089}
Z.~Yin, N.~Cheng, T.~H. Luan, Y.~Song, and W.~Wang, ``{DT}-assisted multi-point
  symbiotic security in space-air-ground integrated networks,'' \emph{IEEE
  Transactions on Information Forensics and Security}, vol.~18, pp. 5721--5734,
  Sep. 2023.

\bibitem{9429703}
Y.~Wu, K.~Zhang, and Y.~Zhang, ``Digital twin networks: A survey,'' \emph{IEEE
  Internet of Things Journal}, vol.~8, no.~18, pp. 13\,789--13\,804, Sep. 2021.

\bibitem{10012285}
Q.~Guo, F.~Tang, T.~K. Rodrigues, and N.~Kato, ``Five disruptive technologies
  in {6G} to support digital twin networks,'' \emph{IEEE Wireless
  Communications}, vol.~31, no.~1, pp. 149--155, Feb. 2024.

\bibitem{9583902}
X.~Lin, J.~Wu, J.~Li, W.~Yang, and M.~Guizani, ``Stochastic digital-twin
  service demand with edge response: An incentive-based congestion control
  approach,'' \emph{IEEE Transactions on Mobile Computing}, vol.~22, no.~4, pp.
  2402--2416, Apr. 2023.

\bibitem{10025677}
S.~Chen, Y.~Wang, D.~Yu, J.~Ren, C.~Xu, and Y.~Zheng, ``Privacy-enhanced
  decentralized federated learning at dynamic edge,'' \emph{IEEE Transactions
  on Computers}, vol.~72, no.~8, pp. 2165--2180, Aug. 2023.

\bibitem{9780389}
J.~Li, W.~Liang, W.~Xu, Z.~Xu, Y.~Li, and X.~Jia, ``Service home identification
  of multiple-source iot applications in edge computing,'' \emph{IEEE
  Transactions on Services Computing}, vol.~16, no.~2, pp. 1417--1430,
  Mar.-Apr. 2023.

\bibitem{Qin2021SemanticCP}
\BIBentryALTinterwordspacing
Z.~Qin, X.~Tao, J.~Lu, and G.~Y. Li, ``Semantic communications: Principles and
  challenges,'' \emph{ArXiv}, 2021. [Online]. Available:
  \url{https://api.semanticscholar.org/CorpusID:245704403}
\BIBentrySTDinterwordspacing

\bibitem{9530497}
G.~Shi, Y.~Xiao, Y.~Li, and X.~Xie, ``From semantic communication to
  semantic-aware networking: Model, architecture, and open problems,''
  \emph{IEEE Communications Magazine}, vol.~59, no.~8, pp. 44--50, Aug. 2021.

\bibitem{9450827}
Z.~Weng and Z.~Qin, ``Semantic communication systems for speech transmission,''
  \emph{IEEE Journal on Selected Areas in Communications}, vol.~39, no.~8, pp.
  2434--2444, Aug. 2021.

\bibitem{9398576}
H.~Xie, Z.~Qin, G.~Y. Li, and B.-H. Juang, ``Deep learning enabled semantic
  communication systems,'' \emph{IEEE Transactions on Signal Processing},
  vol.~69, pp. 2663--2675, Apr. 2021.

\bibitem{AcademicLecture}
Y.~Zhong and R.~Zhang, ``Information ecology and semantic information theory,''
  \emph{Document, Information $\&$ Knowledge}, vol.~0, no.~6, pp. 4--11, Jun.
  2017.

\bibitem{10558819}
F.~Jiang, Y.~Peng, L.~Dong, K.~Wang, K.~Yang, C.~Pan, and X.~You, ``Large {AI}
  model-based semantic communications,'' \emph{IEEE Wireless Communications},
  vol.~31, no.~3, pp. 68--75, Jun. 2024.

\bibitem{10419853}
L.~Wang, W.~Wu, F.~Tian, and H.~Hu, ``Intelligent resource allocation for
  {UAV}-enabled spectrum sharing semantic communication networks,'' in
  \emph{Proc. IEEE 23rd International Conference on Communication Technology
  (ICCT)}, Wuxi, China, 2023, pp. 1359--1363.

\bibitem{9976231}
Q.~Guo, F.~Tang, and N.~Kato, ``Federated reinforcement learning-based resource
  allocation for {D2D}-aided digital twin edge networks in {6G} industrial
  {IoT},'' \emph{IEEE Transactions on Industrial Informatics}, vol.~19, no.~5,
  pp. 7228--7236, May 2023.

\bibitem{10021296}
W.~Liu, B.~Li, W.~Xie, Y.~Dai, and Z.~Fei, ``Energy efficient computation
  offloading in aerial edge networks with multi-agent cooperation,'' \emph{IEEE
  Transactions on Wireless Communications}, vol.~22, no.~9, pp. 5725--5739,
  Sep. 2023.

\bibitem{10234627}
B.~Hazarika, K.~Singh, C.-P. Li, A.~Schmeink, and K.~F. Tsang, ``Radit:
  Resource allocation in digital twin-driven {UAV}-aided internet of vehicle
  networks,'' \emph{IEEE Journal on Selected Areas in Communications}, vol.~41,
  no.~11, pp. 3369--3385, Nov. 2023.

\bibitem{9887906}
Y.~He, M.~Yang, Z.~He, and M.~Guizani, ``Resource allocation based on digital
  twin-enabled federated learning framework in heterogeneous cellular
  network,'' \emph{IEEE Transactions on Vehicular Technology}, vol.~72, no.~1,
  pp. 1149--1158, Jan. 2023.

\bibitem{10335637}
R.~Zhang, Z.~Xie, D.~Yu, W.~Liang, and X.~Cheng, ``Digital twin-assisted
  federated learning service provisioning over mobile edge networks,''
  \emph{IEEE Transactions on Computers}, vol.~73, no.~2, pp. 586--598, Feb.
  2024.

\bibitem{10558825}
J.~Du, T.~Lin, C.~Jiang, Q.~Yang, C.~F. Bader, and Z.~Han, ``Distributed
  foundation models for multi-modal learning in {6G} wireless networks,''
  \emph{IEEE Wireless Communications}, vol.~31, no.~3, pp. 20--30, Jun. 2024.

\bibitem{10419536}
H.~Wang, L.~Wang, and W.~Wu, ``Resource allocation and intelligent trajectory
  optimization for {UAV}-assisted semantic communication system,'' in
  \emph{Proc. IEEE 23rd International Conference on Communication Technology
  (ICCT)}, Wuxi, China, 2023, pp. 1370--1374.

\bibitem{10570867}
W.~C. Ng, H.~Du, W.~Y.~B. Lim, Z.~Xiong, D.~Niyato, and C.~Miao, ``Stochastic
  resource allocation for semantic communication-aided virtual transportation
  networks in the metaverse,'' in \emph{Proc. IEEE Wireless Communications and
  Networking Conference (WCNC)}, Dubai, United Arab Emirates, 2024, pp. 1--6.

\bibitem{10530992}
J.~Tang, J.~Nie, J.~Bai, J.~Xu, S.~Li, Y.~Zhang, and Y.~Yuan, ``U{AV}-assisted
  digital-twin synchronization with tiny-machine-learning-based semantic
  communications,'' \emph{IEEE Internet of Things Journal}, vol.~11, no.~17,
  pp. 28\,437--28\,451, Sep. 2024.

\bibitem{10001594}
L.~Yan, Z.~Qin, R.~Zhang, Y.~Li, and G.~Y. Li, ``Qo{E}-aware resource
  allocation for semantic communication networks,'' in \emph{Proc. IEEE Global
  Communications Conference}, 2022, pp. 3272--3277.

\bibitem{9795902}
B.~Li, Y.~Liu, L.~Tan, H.~Pan, and Y.~Zhang, ``Digital twin assisted task
  offloading for aerial edge computing and networks,'' \emph{IEEE Transactions
  on Vehicular Technology}, vol.~71, no.~10, pp. 10\,863--10\,877, Oct. 2022.

\bibitem{9632276}
Z.~Liang, H.~Chen, Y.~Liu, and F.~Chen, ``Data sensing and offloading in edge
  computing networks: {TDMA} or {NOMA}?'' \emph{IEEE Transactions on Wireless
  Communications}, vol.~21, no.~6, pp. 4497--4508, Jun. 2022.

\bibitem{9420264}
C.~Sun, X.~Wu, X.~Li, Q.~Fan, J.~Wen, and V.~C.~M. Leung, ``Cooperative
  computation offloading for multi-access edge computing in {6G} mobile
  networks via soft actor critic,'' \emph{IEEE Transactions on Network Science
  and Engineering}, vol.~11, no.~6, pp. 5601--5614, Nov.-Dec. 2024.

\bibitem{9681874}
J.~Zhao, L.~Yu, K.~Cai, Y.~Zhu, and Z.~Han, ``{RIS}-aided ground-aerial {NOMA}
  communications: {A} distributionally robust {DRL} approach,'' \emph{IEEE
  Journal on Selected Areas in Communications}, vol.~40, no.~4, pp. 1287--1301,
  Apr. 2022.

\bibitem{10495829}
B.~Li, Z.~Qian, and Z.~Fei, ``{SAC}-based computation offloading for
  reconfigurable intelligent surface-aided mobile edge networks,'' \emph{China
  Communications}, vol.~21, no.~6, pp. 261--270, Jun. 2024.

\end{thebibliography}

\end{document}